\documentclass[conference]{IEEEtran}
\IEEEoverridecommandlockouts
\usepackage{cite,url}
\usepackage{amsmath,amssymb,amsfonts}
\usepackage{algorithmic}
\usepackage{graphicx}
\usepackage{textcomp}
\usepackage{xcolor}
\def\BibTeX{{\rm B\kern-.05em{\sc i\kern-.025em b}\kern-.08em
    T\kern-.1667em\lower.7ex\hbox{E}\kern-.125emX}}
\begin{document}

\title{Blockchain for smart cities improvement: an architecture proposal\\
}

\author{\IEEEauthorblockN{Marco Fiore}
\IEEEauthorblockA{\textit{Department of Electrical and Information Engineering} \\
\textit{Polytechnic University of Bari}\\
Bari, Italy \\
marco.fiore@poliba.it}
\and
\IEEEauthorblockN{Marina Mongiello}
\IEEEauthorblockA{\textit{Department of Electrical and Information Engineering} \\
\textit{Polytechnic University of Bari}\\
Bari, Italy \\
marina.mongiello@poliba.it}
}

\maketitle

\begin{abstract}
The combination between innovative topics and emerging technologies lets researchers define new processes and models. New needs regard the definition of modular and scalable approaches, with society and environment in mind. An important topic to focus on is the smart city one. The use of emerging technologies lets smart cities develop new processes to improve services offered from various actors, either industries or government. Smart cities were born to improve quality of life for citizens. To reach this goal, various approaches have been proposed, but they lack on a common interface to let each stakeholder communicate in a simple and fast way. This paper shows the proposal of an architecture to overcome the actual limitations of smart cities: it uses Blockchain technology as a distributed database to let everyone join the network and feel part of a community. Blockchain can improve processes development for smart cities. Scalability is granted thanks to a context-aware approach: applications do not need to know about the back-end implementation, they just need to adapt to an interface. With Blockchain, it is possible to collect data anonymously to make some statistical analysis, to access public records to ensure security in the city and to guarantee the origin of products and energy.
\end{abstract}

\begin{IEEEkeywords}
Blockchain, smart city, common interface, proposal
\end{IEEEkeywords}

\section{Introduction}
The rise of emerging technologies defines new and improved software processes. The modeling of an architecture is the first step to adapt a process based on specific needs defined by functional and non functional requirements. It is necessary to design some innovative approaches during the definition of the architecture to create a solid and scalable system.
Among various innovative topics in Software Engineering, we focus on smart cities improvement.
The smart city trend is constantly growing as new and emerging technologies help its spreading. A smart city goal is improving the quality of life for citizens, as well as making operations easier and more efficient. To make a smart city desirable, the network should be reliable and with high performances; moreover, privacy and encryption of data should be guaranteed and the concept of trust should be a solid foundation of the entire process.\\
A typical problem in smart cities development is modularity: new applications must be contextualized and developed having that smart city in mind. Actually, there are no standards and guidelines that can adapt to every smart city.

\begin{figure}[t]
\centerline{\includegraphics[width=\linewidth]{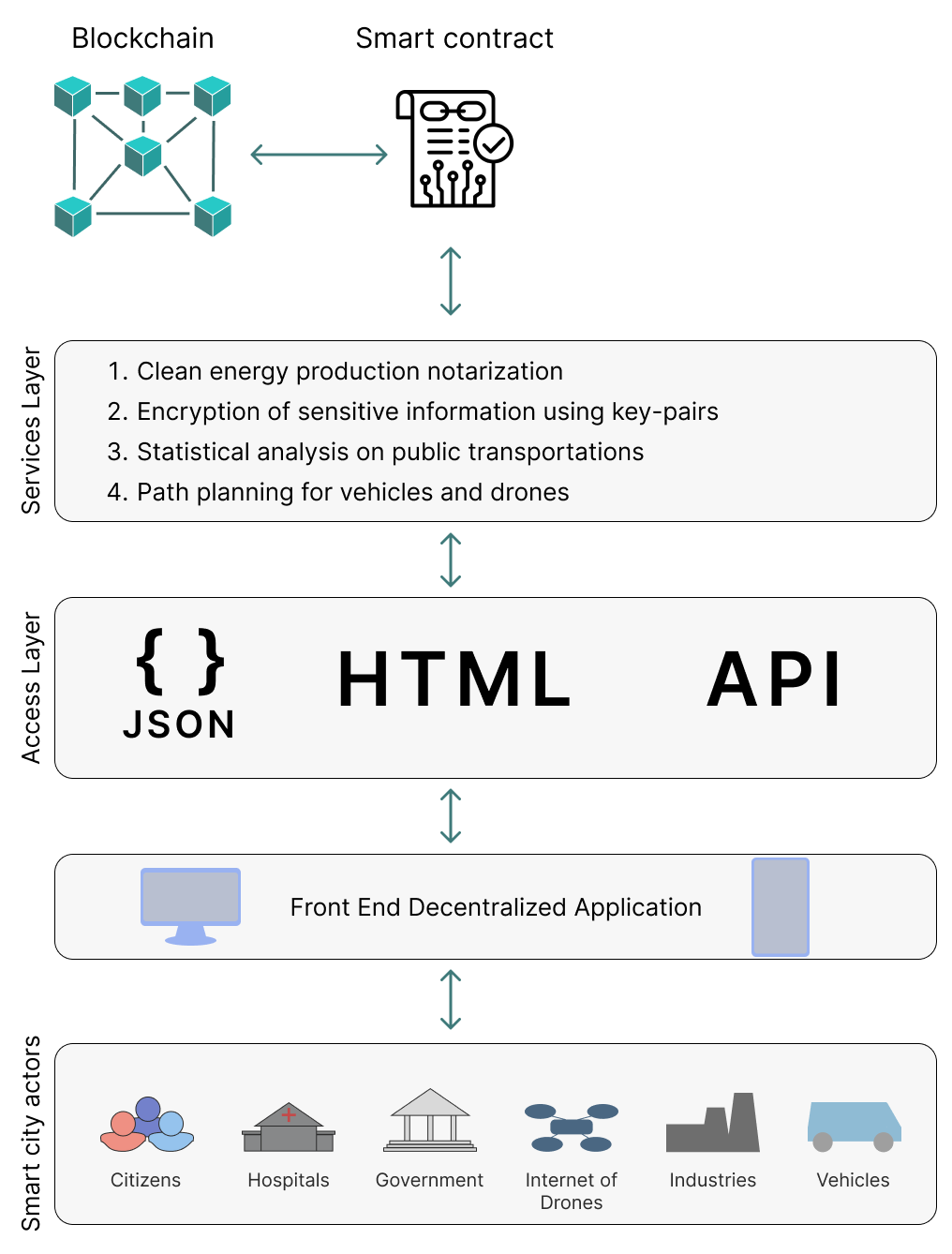}}
\caption{Layers of the system showing a single interface in the access layer}
\label{fig:layers}
\end{figure}

Our proposal is based on the use of Blockchain technology to improve our ability to develop, manage and apply new software and system applications for smart cities.
To illustrate the main aim of this paper, let us consider a sample scenario: suppose we would develop a system that provides smart city services, using a single distributed database enabled for accessing city-related information for citizens. Traditional cities can become smart without using new systems, but simply interfacing with existing ones and with distributed databases used by other smart cities. Hence, as shown in Fig. \ref{fig:layers}, a smart city actor must have a single interface to gather different data and to use the database; this means that the interface and the implementation of an object can vary independently being  separated from one another. The implementation can be realized just once and be compliant to every other smart city that implements the proposed interface.\\

The paper is organized as follows: Section \ref{sec:background} lists some characteristics of Blockchain technology and presents some state of the art analysis to understand what are the benefits and research directions on Blockchain applied to smart cities. Section \ref{sec:vision} discusses the purpose of the paper. Section \ref{sec:architecture} shows the proposal of a scalable architecture that connects each smart city actor to the Blockchain using a common interface, together with some considerations on different applications of such system. Finally, Section \ref{sec:conclusion} concludes the paper.

\section{Blockchain characteristics}\label{sec:background}
Blockchain technology belongs to Distributed Ledger Technologies (DLTs): born in 2008 thanks to Satoshi Nakamoto \cite{nakamoto2008peer}  can be conceived as a distributed database where information are stored in blocks. Each block is connected to the previous and next one thanks to cryptographic (hash) functions. The main Blockchain features are proposed below, to better understand its peculiarities.
\begin{itemize}
    \item Decentralization: all transactions in a Blockchain are inserted and validated through a consensus protocol, then they are replicated between all the nodes participating in the network. In this way, there is no need of a central authority (i.e., a bank from the financial point of view) to maintain all the transactions data.
    \item Immutability: transactions in a Blockchain are stored into blocks. Each block contains its hash and the hash of the previous block, creating a chain that is immutable, because any change on a block will affect all the subsequent blocks. An attacker cannot change the information of a block N without changing all the blocks N+1, N+2,..., M, where M is the total number of blocks in the chain. This change is computationally difficult, so it is not possible to execute it in a short amount of time.
    \item Transparency: the only way to update the ledger is by reaching the consensus by most of the network nodes. All changes are publicly visible: this ensures transparency and security.
    \item Traceability: it becomes easy to trace all transactions in a Blockchain thanks to its immutability and transparency features. In this way, every transaction can be traced down back to its origin.
    \item Trust-less: it is possible to make transaction, in a Blockchain, between unknown parties, even if they do not trust each other. Thanks to the absence of a central authority, it is possible to trust the validity of a transaction without knowing who was involved in it.
\end{itemize}

Different reviews show the role of Blockchain in smart cities, with some focus on smart healthcare, smart transportation, supply chains \cite{xie2019survey}. They also underline the combination of Blockchain and other technologies such as Internet of Things and Machine Learning \cite{rejeb2022blockchain}. Blockchain can help smart cities in being more sustainable thanks to its peculiarities: a) it is immutable, so every information added to the chain cannot be modified, b) it is anonymous, meaning that everyone can join the network without worrying about privacy, c) it is trustable even if people don't know each other.

Authors of paper \cite{ullah2021conceptual} focus their research on Blockchain smart contracts in smart real estate. They propose a conceptual framework for the adoption of such topic in smart cities. The real estate process becomes more immersive and user-friendly, in line with Industry 4.0 requirements.

Blockchain can help smart cities development \cite{shari2022state} both from performance and security sides. The distributed nature of this technology makes architectures more scalable and with less point of failures: as soon as one node is active, the entire network is up. Data sharing takes advantage of this approach: education, healthcare, buildings can communicate using a single common interface. Artificial Intelligence intervenes in data management and analysis \cite{sharma2021sustainable}: deep learning techniques can enrich the green energy production \cite{godahewa2022generative, olu2022building}, while neural networks can improve road management \cite{saleem2022smart}.

Traceability characteristic of Blockchain is helpful for waste management \cite{ahmad2021blockchain} thanks to notarized documentation, compliance with laws and fleet management. It is also useful with respect to public emergency services \cite{kumar2022best}: it can help security workers to manage different anomalies, from fires to crimes.

A state-of-the-art summary is shown in Table \ref{tab:sota}. The analysis of such publications raises some open challenges:
a) Sustainability is an important aspect in the topic of Blockchain applied to smart cities. It is the furthest research topic from most of the analyses \cite{rejeb2022blockchain}. b) There is the need of a single interface to the Blockchain, to create a bridge between different actors in the smart city and a single, common distributed database. c) Security and privacy should be underlined\cite{ghazal2022securing}: Blockchain preserves privacy and ensures that only authorized nodes can access sensitive information. d) Costs to deploy a complete Blockchain network in a smart city are not yet known. It is difficult to perform a cost prediction in the deployment of a Blockchain in a smart city \cite{xie2019survey}. e) Regulations are needed to correctly share information: smart contracts can come in hand in this topic.

\begin{table}[htbp]
    \caption{State of the art analysis}
    \begin{center}
        \begin{tabular}{|p{2.5cm}|p{1.5cm}|p{3cm}|}
            \hline
            \textbf{Paper} & \textbf{Publication year} & \textbf{Topic} \\
            \hline
            xie2019survey\cite{xie2019survey} & 2019 & Survey on the literature involving Blockchain technology applied to smart cities\\
            \hline
            rejeb2022blockchain\cite{rejeb2022blockchain} & 2022 & Trends and research directions for Blockchain applied to smart cities\\
            \hline
            ullah2021conceptual\cite{ullah2021conceptual} & 2021 & Usage of smart contracts in smart real estate environment\\
            \hline
            shari2022state\cite{shari2022state} & 2022 & Survey of Blockchain applications for data management in smart cities\\
            \hline
            sharma2021sustainable\cite{sharma2021sustainable} & 2021 & Integration of Blockchain and Artificial Intelligence for sustainable smart cities\\
            \hline
            ahmad2021blockchain\cite{ahmad2021blockchain} & 2021 & Usage of Blockchain for waste management in smart cities\\
            \hline
            kumar2022best\cite{kumar2022best} & 2022 & Protection of life and properties from fire damage in smart cities using Blockchain\\
            \hline
            ghazal2022securing\cite{ghazal2022securing} & 2022 & Protection of smart cities using Blockchain as a distributed database\\
            \hline
            \end{tabular}
        \label{tab:sota}
    \end{center}
\end{table}
\section{Our vision}\label{sec:vision}
We envision a scenario in which Blockchain is the foundation of smart cities processes. Each process can be easily added to the system thanks to a common interface that embraces every aspect of the city. Information can be exchanged using JSON format, so the communication between a front end decentralized application and the Blockchain is context-aware. Blockchain technology has the potential to play an important role in the development of smart cities. It can provide multiple advantages in many topics:
\begin{itemize}
    \item Supply chain management: Blockchain can be used to track goods and materials through a supply chain, thus increasing transparency and reducing the risk of fraud.
    \item Sustainability: Blockchain can be used to manage and track the use of renewable energy. A sustainable smart city can be obtained if actors reduce their carbon footprint and promote green approaches.
    \item Authentication and identification: Blockchain can be used to verify identities in a secure and decentralized way, making it easier for citizens to access services and participate in civic life.
    \item Public records: Blockchain can be used to store and manage public records, such as property titles or licenses.
    \item Transportation: Blockchain can be used to manage and track the use of public transportation, helping cities optimize their transportation systems and reduce congestion. Transports side, Blockchain can be used to gather information to improve paths, waiting times and overall services.
\end{itemize}

Overall, the role of Blockchain in smart cities management is to improve different aspects, from sustainability (i.e., notarization of clean energy production) to hijacking avoidance (i.e., guaranteeing the path of a bus or a taxi, making statistical analysis for public transports, identifying passengers). The ultimate goal is to improve quality of life for citizens.

\subsection{Clean energy production}
Smart buildings must be energy efficient and incorporate clean energy production technologies. The ways to accomplish this goal are different: a) solar panels can be installed on the roof of a building to capture sunlight and generate electricity, b) wind turbines can convert wind speed to electricity, c) storage systems, such as batteries, can store excess clean energy produced during low demand times. Blockchain technology can support the production of clean energy in multiple ways: a) it can track and verify the energy production, to ensure that a building is sustainable; b) it can help the trading of energy, notarizing transactions between a building with enough energy in its storage and a building with less energy than requested; c) it can help people understand if a building is really sustainable and green (i.e., showing a building carbon footprint), thus letting people choose and prefer smarter and more efficient buildings.

\subsection{Encryption of sensitive information}
Citizens side, information should be encrypted to ensure privacy and anonimity. The encryption process can be both symmetric or asymmetric. In this proposal, we follow an asymmetric key encryption scheme, that takes advantage of the key-pair already present in every Blockchain architecture. In this way, everyone can encrypt any kind of message using the recipient public key, thus guaranteeing that only the recipient can decrypt the message using his or her own private key.

\begin{figure}[htbp]
\centerline{\includegraphics[width=\linewidth]{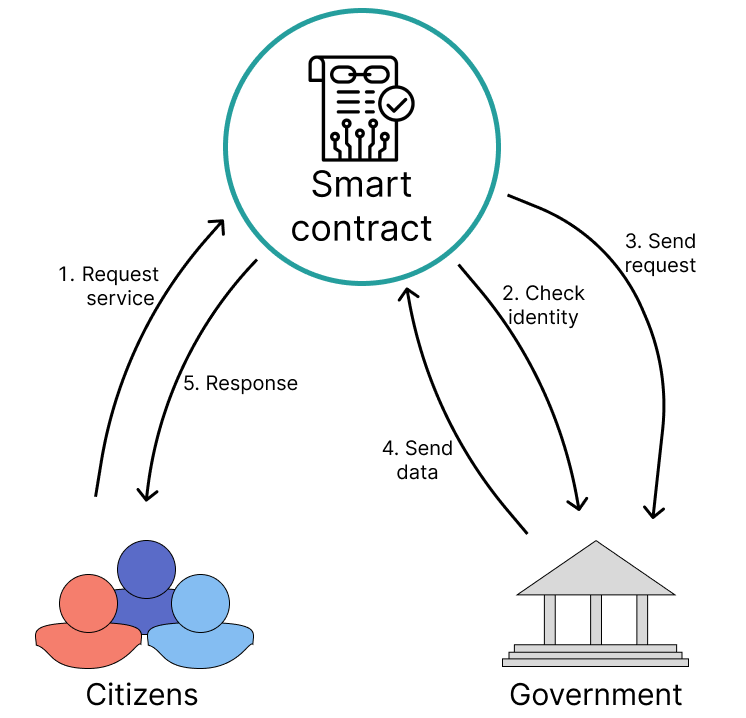}}
\caption{Sample process for a citizen requesting a service to government institutions}
\label{fig:authentication}
\end{figure}

In the case of public services requests, it is possible to use smart contracts to make the process secure and transparent. The authentication and request process is shown in Fig. \ref{fig:authentication}.
\begin{enumerate}
    \item The citizen requests a service to the institution (i.e., a certificate of residence). The request is managed by a smart contract. It is also possible to directly upload documents to the InterPlanetary File System (IPFS) \cite{shi2022secure}, but due to lack of regulations and laws, we decided to let institutions keep sensitive documents.
    \item The smart contract, together with the institution, authenticates the citizen and ensures that the requested certificate is obtainable.
    \item The smart contract requests the document to the institution.
    \item The institution gives back the requested service using the same smart contract.
    \item The citizen receives the requested service or document. The process, thanks to smart contracts intervention, is transparent, secure and fast.
\end{enumerate}

\subsection{Statistical analysis}
The use of a distributed database such as Blockchain lets people read public records, that are stored in the chain and accessible to anyone. These data are stored anonymously, meaning that any information can be related to a public key (wallet), but there is no way to link that wallet to a person. In this way, data can be used to make some statistical analysis to understand how to improve services offered to citizens. Public transports can easily understand if there is a lack on the offer and improve it, knowing exactly where to act.

\subsection{Path planning}
The process of determining one optimal route for a vehicle to travel from one location to another is defined as path planning. This approach can be used to avoid traffic congestion in smart cities \cite{lin2020spatiotemporal} or to quickly intervene in case of disasters \cite{qadir2021addressing}. Path planning can be used for vehicles, drones and people. With Blockchain technology, it is possible to avoid hijacking: in the Internet of Drones (IoD) field, various approaches have been proposed \cite{abualigah2021applications, allouch2021utm, singh2022referenced}. They all share a common point of view: everytime a drone approaches a new Point of Interest (PoI), it writes a new information on Blockchain to notarize its position. In this way, every attempt of hijacking can be identified in short time. The same approach can be considered for Autonomous Guided Vehicles (AGVs) in a smart city: AGVs can read from the Blockchain where they have to go, then they can create an optimal path and notarize the time of arrival. These information can further be used for statistical analysis, as underlined before.

\section{Proposed architecture}\label{sec:architecture}
\begin{figure}[htbp]
\centerline{\includegraphics[width=\linewidth]{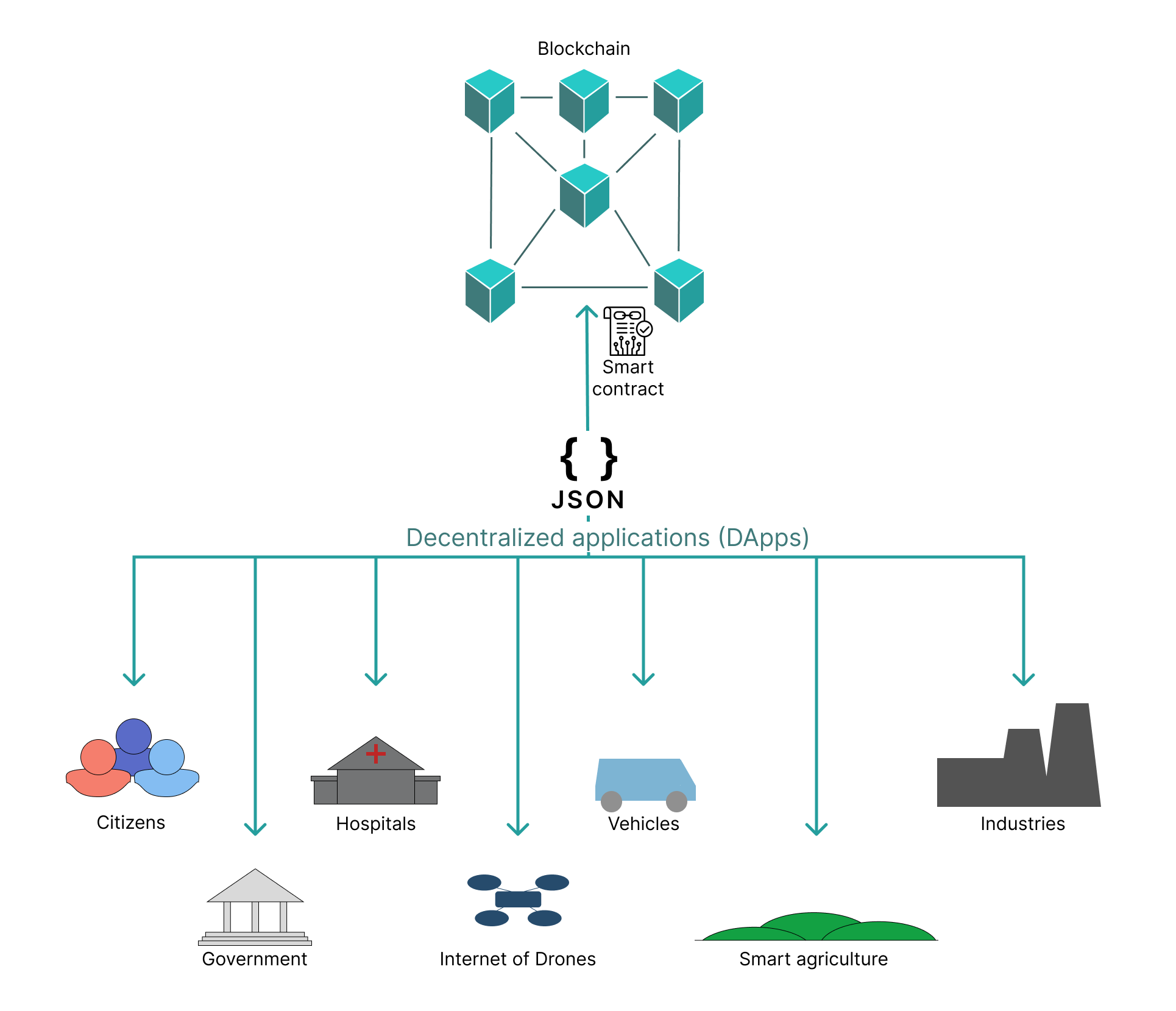}}
\caption{Architecture of the proposed Blockchain-based smart city}
\label{fig:architecture}
\end{figure}

Our contribution regards the design of an architecture where every actor in a smart city can benefit from using Blockchain as a back end of the system. The main goal is the development of a common interface to communicate with the database, so everyone can join the network in a secure and fast way. Smart contracts can receive any kind of data in a JSON format: new actors just need to upload JSON-formatted information. Data are then managed by the contract, that gathers them and converts them into value, thus uploading them in the Blockchain. An architecture showing different actors is proposed in Fig. \ref{fig:architecture}.

In our scenario, each actor uses a Decentralized Application (DApp) to connect to the Blockchain. DApps are designed to be distributed and to run on multiple nodes, rather than being controlled by a single entity. Some applications of such architecture can be summed up as follows:
\begin{itemize}
    \item \textit{A. Clean energy production.} To reach sustainability purposes, clean energy production can be notarized in Blockchain. Everyone can ensure that the energy used in a building comes from renewable sources.
    \item \textit{B. Encryption of sensitive information.} Sensitive information can be encrypted using nodes key-pair. Autonomous shared vehicles (i.e., taxies) can use this sign to authenticate passengers and ensure that only the passenger who payed for the ride can use that vehicle.
    \item \textit{C. Statistical analysis.} Public transportation can make statistical analysis (i.e., preferred destination, waiting times, etc.) to improve the offer, still guaranteeing anonymity.
    \item \textit{D. Path planning.} Path planning is possible to avoid hijacking. In the Internet of Drones topic, this can be a useful approach to ensure that the path followed by a drone is correct and there is no tampering \cite{liao2021securing, singh2022referenced}.
\end{itemize}

Besides, every actor in the smart city can feel as part of a community, easily accessing any public information in the Blockchain and exchanging messages with other actors in a transparent way.

The architecture respects requirements for building a system process with modularity and scalability in mind, thus ensuring high performances and reliability that are guaranteed by the presence of Blockchain.\\

\begin{figure}[htbp]
\centerline{\includegraphics[width=.9\linewidth]{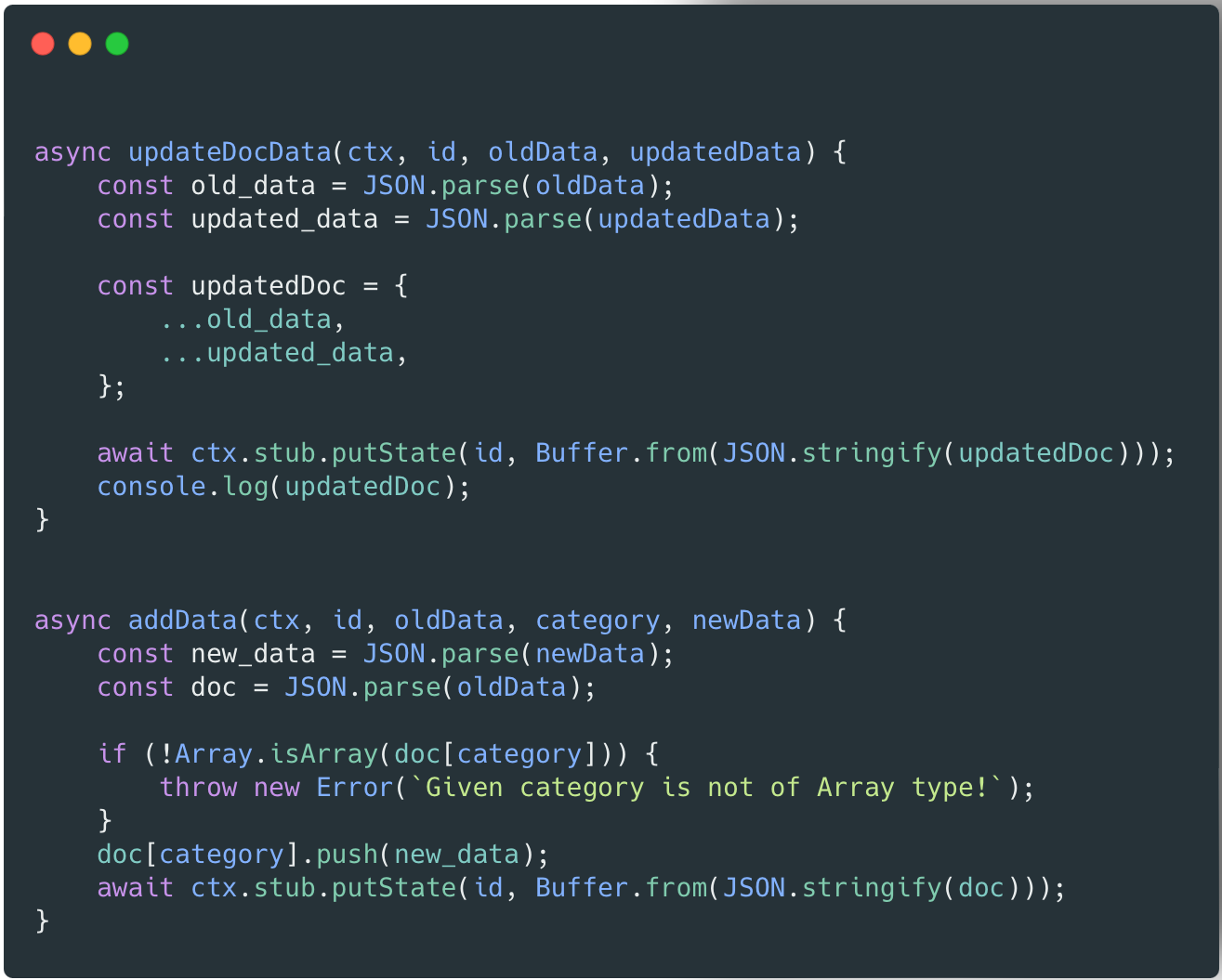}}
\caption{Context-aware smart contract supporting JSON-formatted information}
\label{fig:api}
\end{figure}

\begin{figure}[htbp]
\centerline{\includegraphics[width=\linewidth]{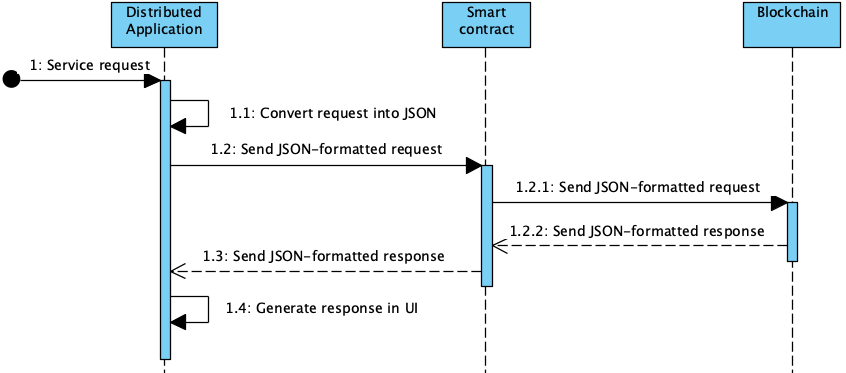}}
\caption{Sequence diagram for a smart city service request}
\label{fig:sequence}
\end{figure}

A prototype is being developed to show the advantages of adopting a single access layer. To upload JSON information to the Blockchain, some context-aware smart contracts are designed, as proposed in Fig. \ref{fig:api}. These smart contracts take the input, make some checks on the correctness of data and then upload them to the Blockchain. Data are accessible by smart city actors; the retrieval process gives as output a JSON object. The specific front end distributed application can manage the JSON output to show the information requested by the user. The described process is shown in Fig. \ref{fig:sequence}: steps from 1.2 to 1.3 are independent from the front end distributed application.

\section{Conclusion}\label{sec:conclusion}
In this paper we propose an architecture for applying Blockchain technology in smart cities. Thanks to its characteristics, Blockchain guarantees anonymity, optional encryption of data and traceability of goods, clean energy production and vehicles. We show how different stakeholders can benefit from using this architecture, then we make some considerations on sample workflows (i.e., the process for a citizen who requests a service or document to government institutions) and processes improvement. The context-aware approach grants scalability and lets applications interface in the same way to every service proposed in the city.
Future developments regard the development of a complete simulator using smart contracts to implement some smart city services, to show how each actor can interact, exchange data and access public records.

\bibliographystyle{ieeetr}
\bibliography{refs}

\begin{thebibliography}{10}

\bibitem{nakamoto2008peer}
S.~Nakamoto and A.~Bitcoin, ``A peer-to-peer electronic cash system,'' {\em
  Bitcoin.--URL: https://bitcoin. org/bitcoin. pdf}, vol.~4, p.~2, 2008.

\bibitem{xie2019survey}
J.~Xie, H.~Tang, T.~Huang, F.~R. Yu, R.~Xie, J.~Liu, and Y.~Liu, ``A survey of
  blockchain technology applied to smart cities: Research issues and
  challenges,'' {\em IEEE Communications Surveys \& Tutorials}, vol.~21, no.~3,
  pp.~2794--2830, 2019.

\bibitem{rejeb2022blockchain}
A.~Rejeb, K.~Rejeb, S.~J. Simske, and J.~G. Keogh, ``Blockchain technology in
  the smart city: A bibliometric review,'' {\em Quality \& Quantity}, vol.~56,
  no.~5, pp.~2875--2906, 2022.

\bibitem{ullah2021conceptual}
F.~Ullah and F.~Al-Turjman, ``A conceptual framework for blockchain smart
  contract adoption to manage real estate deals in smart cities,'' {\em Neural
  Computing and Applications}, pp.~1--22, 2021.

\bibitem{shari2022state}
N.~F.~M. Shari and A.~Malip, ``State-of-the-art solutions of blockchain
  technology for data dissemination in smart cities: A comprehensive review,''
  {\em Computer Communications}, 2022.

\bibitem{sharma2021sustainable}
A.~Sharma, E.~Podoplelova, G.~Shapovalov, A.~Tselykh, and A.~Tselykh,
  ``Sustainable smart cities: convergence of artificial intelligence and
  blockchain,'' {\em Sustainability}, vol.~13, no.~23, p.~13076, 2021.

\bibitem{godahewa2022generative}
R.~Godahewa, C.~Deng, A.~Prouzeau, and C.~Bergmeir, ``A generative deep
  learning framework across time series to optimize the energy consumption of
  air conditioning systems,'' {\em IEEE Access}, vol.~10, pp.~6842--6855, 2022.

\bibitem{olu2022building}
R.~Olu-Ajayi, H.~Alaka, I.~Sulaimon, F.~Sunmola, and S.~Ajayi, ``Building
  energy consumption prediction for residential buildings using deep learning
  and other machine learning techniques,'' {\em Journal of Building
  Engineering}, vol.~45, p.~103406, 2022.

\bibitem{saleem2022smart}
M.~Saleem, S.~Abbas, T.~M. Ghazal, M.~A. Khan, N.~Sahawneh, and M.~Ahmad,
  ``Smart cities: Fusion-based intelligent traffic congestion control system
  for vehicular networks using machine learning techniques,'' {\em Egyptian
  Informatics Journal}, 2022.

\bibitem{ahmad2021blockchain}
R.~W. Ahmad, K.~Salah, R.~Jayaraman, I.~Yaqoob, and M.~Omar, ``Blockchain for
  waste management in smart cities: A survey,'' {\em IEEE Access}, vol.~9,
  pp.~131520--131541, 2021.

\bibitem{kumar2022best}
S.~Kumar, R.~S. Rathore, M.~Mahmud, O.~Kaiwartya, and J.~Lloret,
  ``Best—blockchain-enabled secure and trusted public emergency services for
  smart cities environment,'' {\em Sensors}, vol.~22, no.~15, p.~5733, 2022.

\bibitem{ghazal2022securing}
T.~M. Ghazal, M.~K. Hasan, H.~M. Alzoubi, M.~Al~Hmmadi, N.~A. Al-Dmour,
  S.~Islam, R.~Kamran, and B.~Mago, ``Securing smart cities using blockchain
  technology,'' in {\em 2022 1st International Conference on AI in
  Cybersecurity (ICAIC)}, pp.~1--4, IEEE, 2022.

\bibitem{shi2022secure}
D.~Shi, C.~Cao, and J.~Ye, ``Secure government data sharing based on blockchain
  and attribute-based encryption,'' in {\em International Symposium on Security
  and Privacy in Social Networks and Big Data}, pp.~324--338, Springer, 2022.

\bibitem{lin2020spatiotemporal}
C.~Lin, G.~Han, J.~Du, T.~Xu, L.~Shu, and Z.~Lv, ``Spatiotemporal
  congestion-aware path planning toward intelligent transportation systems in
  software-defined smart city iot,'' {\em IEEE Internet of Things Journal},
  vol.~7, no.~9, pp.~8012--8024, 2020.

\bibitem{qadir2021addressing}
Z.~Qadir, F.~Ullah, H.~S. Munawar, and F.~Al-Turjman, ``Addressing disasters in
  smart cities through uavs path planning and 5g communications: A systematic
  review,'' {\em Computer Communications}, vol.~168, pp.~114--135, 2021.

\bibitem{abualigah2021applications}
L.~Abualigah, A.~Diabat, P.~Sumari, and A.~H. Gandomi, ``Applications,
  deployments, and integration of internet of drones (iod): a review,'' {\em
  IEEE Sensors Journal}, 2021.

\bibitem{allouch2021utm}
A.~Allouch, O.~Cheikhrouhou, A.~Koub{\^a}a, K.~Toumi, M.~Khalgui, and
  T.~Nguyen~Gia, ``Utm-chain: blockchain-based secure unmanned traffic
  management for internet of drones,'' {\em Sensors}, vol.~21, no.~9, p.~3049,
  2021.

\bibitem{singh2022referenced}
M.~P. Singh, A.~Singh, G.~S. Aujla, R.~S. Bali, and A.~Jindal, ``Referenced
  blockchain approach for road traffic monitoring in a smart city using
  internet of drones,'' in {\em ICC 2022-IEEE International Conference on
  Communications}, pp.~1--6, IEEE, 2022.

\bibitem{liao2021securing}
S.~Liao, J.~Wu, J.~Li, A.~K. Bashir, and W.~Yang, ``Securing collaborative
  environment monitoring in smart cities using blockchain enabled
  software-defined internet of drones,'' {\em IEEE Internet of Things
  Magazine}, vol.~4, no.~1, pp.~12--18, 2021.

\end{thebibliography}

\end{document}